\documentclass[11pt]{article}
\usepackage{rotating}

\usepackage{lscape}

\usepackage{graphicx}
\begin{document}
\title{\bf Peculiarities of Abundances of Selected Elements
in Metal-Rich Field RR~Lyrae Stars}

\author{{M.\,L.~Gozha, V.\,A.~Marsakov, V.\,V.~Koval'}\\
{Southern Federal University, Rostov-on-Don, Russia}\\
{e-mail:  gozha\_marina@mail.ru, marsakov@sfedu.ru, vvkoval@sfedu.ru}}
\date{accepted \ 2020, Astrophysical Bulletin, Vol. 75, No. 3, pp. 311-319}

\maketitle

\begin {abstract}

We use the data of our extended catalog of spectroscopic 
determinations of elemental abundances in
the atmospheres of Galactic-field RR Lyrae type variables to show 
that metal-rich RR~Lyraes ($\rm{[Fe/H]} > -1.0$)
have anomalous abundances of some elements. In particular, the 
relative abundances of scandium, titanium,
and yttrium in metal-rich RR~Lyrae type variables are lower than 
the corresponding abundances in field stars of
similar metallicity beyond the errors. We discuss the errors of 
the determination of the abundances of the above
elements and point out the fact that no europium, zirconium, 
and lanthanum abundance determinations are
available for metal-rich RR~Lyrae type variables. We also 
analyze various possible causes of the observed peculiarities
of the chemical composition of metal-rich RR~Lyrae type variables.

\end{abstract}


\maketitle

\section{Introduction}

RR~Lyrae type stars are radially pulsating A--F-type
low-mass variables populating the instability strip
of the horizontal branch. Metal abundances of Galactic-field 
RR~Lyraes vary over a wide range. Unlike
globular-cluster RR~Lyraes, some of Galactic-field
RR~Lyraes have high metallicities and kinematic properties
typical of disk subsystems. RR~Lyrae type variables
are stars at the late evolutionary stage. However,
studies of RR~Lyraes show that the abundances of
most of the heavy elements in their surface layers
remain unchanged (see, e.\,g., Clementini et\,al.,~1995).
Hence the [el/Fe] ratios can serve as indicators of the
chemical composition of the matter from which these
stars formed.

In our earlier papers (Marsakov et\,al.,~2018a, b) we
used our catalog containing the data about the positions,
velocities, and metallicities for 415 field RR~Lyrae 
type variables and relative abundances [el/Fe] of
12 chemical elements in one hundred RR Lyrae type
variables to show that metal-rich ($\rm{[Fe/H]} > -1.0$) RR
Lyraes with low relative abundances of $\alpha$-elements
($\rm{[\alpha/Fe]} < 0.2$) in our Galaxy have kinematics that is
typical of the youngest subsystem -- the thin Galactic
disk. Low residual velocities of these variables indicate
that such stars must be younger than the oldest stars of
this subsystem, i.\,e., their ages should be smaller than
9~Gyr (see, e.\,g., Bensby et\,al.,~2014). However, our
estimates based on Dartmuth evolutionary tracks
(Marsakov et\,al.~ 2019a, b) have shown that the masses
of these stars are rather small ($0.51-0.60 M_{\odot}$), and,
according to current views, their initial masses should
have been of about $0.7-0.8 M_{\odot}$ (see Taam et\,al.,~1976,
and references therein). The evolutionary time scale of
such low-mass stars is longer than the age of the thin
disk. We suggested in our paper (Marsakov et\,al.,
~2019a) that the presence of such young and metal-rich
RR~Lyraes may be due to high helium abundances
of their progenitors.

In our earlier studies (Marsakov et\,al.,~2018a, b,
2020) we investigated the dependencies of relative
abundances of magnesium, silicon, calcium, and titanium
(representing $\alpha$-elements) on [Fe/H] in F--G-type
dwarfs and field RR~Lyraes. On the whole, the
metallicity dependencies of the abundances of $\alpha$-elements
in RR Lyraes practically reproduce the corresponding
dependencies for field dwarfs. However, at
$\rm{[Fe/H]} > -1.0$ the relative titanium abundances in
most of the RR~Lyraes are significantly lower than in
field dwarfs (note that such RR~Lyrae have subsolar
[Ti/Fe] ratios). The authors of the original studies we
cite already pointed out the anomalous abundances of
some elements. In particular, Chadid et\,al.~(2017);
Clementini et\,al. (1995); Liu et\,al.~(2013) also tried to
explain the lower relative abundances not only of titanium,
but also of scandium and yttrium in RR Lyraes
with $\rm{[Fe/H]} > -1.0$ compared to stationary stars of the
same metallicity.

Thus Clementini et\,al. (1995) point out that the
anomalously low scandium and yttrium abundances
that they found in two metal-rich RR~Lyraes of their
sample. The above authors could not explain such
results by a combination of errors of atmospheric
parameters or attribute them to nucleosynthesis processes.
They suggested that such low abundances of
these elements could have resulted from superionization
caused by photons emitted in Lyman lines, which
are induced by shocks in the pulsating atmospheres of
RR~Lyraes because stationary stars with the atmospheric
parameters like those of RR~Lyrae variables
exhibit no scandium or yttrium deficit. Underestimation
of the effect of superionization in LTE model
atmospheres may result in underestimated abundances
of these elements. However, the above authors
and Liu et\,al.~(2013) considered such explanation
unpersuasive because of the lack of a similar effect in
RR~Lyraes with $\rm{[Fe/H]} < -1.0$.

In contrast to the above explanation, Liu et\,al.~(2013) 
estimate the possible effect of the fact that
scandium lines are subject to hyperfine splitting and
found this effect to be insignificant. They argue that
anomalously low relative abundances of Sc and Y in
metal-rich RR~Lyrae type stars cannot be explained
and there fire they cannot be used for analyzing the
chemical evolution of our Galaxy. The above authors
suggest that given that the [Sc/Fe] and [Y/Fe] ratios
are determined using lines of ionized atoms, which are
sensitive to surface gravity, the deviations can be due to
differences of log\,g between RR~Lyrae type variables
and dwarfs. However, our analysis of numerous determinations
of surface gravity in metal-rich RR~Lyraes
from high-resolution spectra showed that these
parameters are confined within the log\,g = $2.5-3.0$
interval, i.\,e., they are more or less the same as in red
giants (see, e.\,g., Gozha et\,al.,~2019; 
Marsakov et\,al.,~2019a). Moreover, the surface gravities of other 
metal-rich variables -- Cepheids —- are even lower than those
of RR~Lyraes (e.\,g., Luck,~2018), whereas the [el/Fe]
ratios of the elements considered (especially those of
scandium and yttrium) are, on the average, somewhat
higher in Cepheids than in field dwarfs and giants and
than in metal-rich RR~Lyraes (see below).

Chadid et\,al.~(2017) argue that the nature of metal-rich
RR~Lyrae type variables is unique despite the fact
that both metal-poor and metal-rich RR~Lyraes are
helium-core burning horizontal-branch stars. They
believed that the dynamics and structure of the atmospheres
of these stars differ. The differences, which are
most conspicuous in RRab type variables, show up in
the strength and localization of shocks. Note that in
metal-rich RR~Lyraes the strength of main shock is
higher in photospheres, whereas in more metal-poor
RR~Lyraes it is higher in the upper layers of the atmospheres.
The above authors believe that the differences
in the mechanisms of shock propagation are due to the
differences between the atmospheric parameters of
these stars, because in their sample metal-rich RR
Lyraes have higher effective temperatures and surface
gravities than more metal-poor stars. However,
although the $T_{eff}$ and log\,g values that we inferred for
one hundred RR~Lyraes (Marsakov et\,al.,~2019a) are,
on the average, slightly higher for metal-rich stars,
they nevertheless agree within the quoted errors. Note
that all metal-rich RR~Lyraes are located inside broad
intervals of the corresponding parameters spanned by
metal-poor RR~Lyraes.

Given the inconclusive nature of the hypotheses
suggested by different authors to explain the peculiarities
of the abundances of some chemical elements in
Galactic field RR~Lyrae type variables, here we critically
review the proposed arguments and analyze in
detail the relative abundances in metal-rich ($\rm{[Fe/H]} >
-1.0$) field RR~Lyraes of the elements that exhibit
anomalies. Our careful preliminary analysis of all
available determinations of elemental abundances
allowed us to identify the elements that clearly exhibit
anomalous relative abundances in metal-rich RR~Lyraes. 
We found these to be scandium, titanium,
yttrium, as well as europium, zirconium, and lanthanum --
the latter three have never been found in any
metal-rich RR~Lyrae type variable.


\section {FORMATION OF CHEMICAL ELEMENTS}

According to current theoretical views, isotopes of
chemical elements form in certain nuclear synthesis
reactions in stars of various masses. Let is recall how
and where the above chemical elements are synthesized.

Scandium is an iron-peak element with odd number
of protons. It is believed to be synthesized in massive
stars during explosive burning of oxygen and silicon
as the radioactive progenitor of $^{45}\rm{Ti}$, and also in
the process of neon burning. It also synthesized in the
envelopes of the same stars in the weak component of
the slow neutron capture processes (the $s$-process)
during the helium and carbon envelope burning
(Ernandes et\,al.,~2018; Limongi and Chieffi,~2003;
Woosley and Weaver,~1995).

Titanium is classified an $\alpha$-element, but in some
cases it is referred to as an iron-group element (e.\,g.,
Sneden et\,al.,~2016). This element is synthesized
during explosive burning of silicon and oxygen mostly
by massive ($M > 10 M_{\odot}$) type II supernovas and, in
small amounts, by SN Ia (Thielemann et\,al.,~2002;
Tsujimoto et\,al.,~1995; Woosley and Weaver,~1995). It
has long been observed that relative titanium abundances
depend on metallicity similarly to the relative
abundances of other $\alpha$-elements.

Unfortunately, current theoretical models are incapable
to reproduce the observed trends for [Sc/Fe] and
[Ti/Fe], and therefore the nucleosynthesis problem
for scandium and titanium remains unsolved (see
Mishenina et\,al.,~2017).

Yttrium is synthesized in slow neutron capture processes.
$S$-process elements form in low- and intermediate-
mass stars ($1.5 - 8 M_{\odot}$) during the double-shell
burning stage (the main component of the $s$-process)
as a result of thermal pulsations in the envelopes of
asymptotic-branch giants (AGB stars), and during the
ejection of the envelope they are released to the interstellar
space (Gallino et\,al.,~1998). A certain fraction of
these elements are synthesized in the cores of massive
($M \geq 8M_{\odot}$) stars during the hydrostatic stage of the
helium core burning (the weak component of the $s$-process).
The number of neutrons required for $s$-processes
is synthesized in AGB star in the process of the
reactions $^{13}C(\alpha, n)^{16}O$ (Bisterzo et\,al.,~2011; Gallino
et\,al.,~1998), or, in the interiors of massive stars these
elements are synthesized in reactions $^{22}Ne(\alpha, n)^{25}Mg$
(Pignatari et\,al.,~2010; Woosley et\,al.,~1994). The main
component provides about $70-90\,\%$ of the yttrium
abundance in the Sun (e.\,g., Arlandini et\,al.,~1999;
Travaglio et\,al.,~2004). Yttrium can also be synthesized
in the $r$-process.

Zirconium and lanthanum owe their origin to the
$s$-process.

Europium is a typical rapid neutron capture element
(the $r$-process). Ninety-four percent of the europium
abundance in the Sun is provided by the $r$-process
(Bisterzo et\,al.,~2017). It is believed that the synthesis
of the bulk amount of the atoms of rapid
neutron capture elements occurs during explosions of
type-II supernovas with masses $8-10M_{\odot}$ (Marsakov
et\,al.,~2019a). Although particular mechanisms of the
production of $r$-process elements have not yet been
entirely understood, they in any case are associated
with the final stages of the evolution of short-lived
massive stars (Thielemann et\,al.,~2011).

\section {INITIAL DATA} 

For our analysis of the behavior of the above chemical
elements in Galactic-field RR~Lyrae type stars we used
the spectroscopic determinations of metallicity and relative
abundances of titanium, yttrium, zirconium, lanthanum,
and europium for one hundred field RR~Lyraes
from our compiled catalog described in Marsakov et\,al.~(2018a). 
The catalog is available at http://vizier.ustrasbg.
fr/viz-bin/VizieR?-source=J/AZh/95/54. The
[Fe/H] and [el/Fe] relative abundances in this catalog
are compiled from 25 publications, and the case of two
or more determinations by different authors the
weighted averages are computed and reduced to the
common solar abundance (see Gozha et\,al.,~2018).

We added to these data relative scandium abundances
[Sc/Fe] for 77 RR~Lyraes adopted from
15 papers published from 1995 through 2017. The original
studies used high-resolution spectra, which were
analyzed in terms of LTE approximation. The scandium
abundances for 31 stars were determined in several
(two to four) papers. In those cases, we computed
the weighted averages with the coefficients inversely
proportional to the errors quoted by the authors. The
authors of several papers did not provide errors of their
[Sc/Fe] determinations, and in those cases we set the
uncertainties equal to 0.1~dex when computing the
weighted averages. To homogenize the data, we
reduced the [Sc/Fe] determinations to the solar abundance
from Asplund et\,al.~(2009). Note that as a result
we obtained an impressive list of relative [Sc/Fe]
abundances in the atmospheres of RR~Lyraes and we
found no other published list of comparable size.

Table 1 lists the compiled scandium abundances
and the averages that we computed. Column 1 gives
the name of the star. Column 2 gives the [Fe/H] values
for 68 RR~Lyraes of our catalog (Marsakov et\,al.,~2018a) 
and for nine RR~Lyraes from Sneden et\,al.~(2017), 
six of which are not included into our catalog
and three other stars lack chemical-composition
determinations in Marsakov et \,al.~(2018a): these 
RR~Lyraes are marked by the asterisk in the Table. The
next column gives the final [Sc/Fe] values. The last
column gives the references to the [Sc/Fe] data
sources. The literature used to determine [Fe/H],
[Ti/Fe], and [Y/Fe] is listed in catalog (Marsakov
et\,al.,~2018a).

For comparison, we used several samples of field
stars with known metallicities and relative scandium,
titanium, and yttrium abundances. We adopted the
data for 7066 nearby dwarfs, subgiants, and turn-off
stars of disk subsystems from Buder et\,al.~(2019). Catalog
by Reddy et\,al.~(2006) served as the source of
[Fe/H] and [Sc/Fe] for 171 F--G-type dwarfs located
within 150 pc from the Sun, which in most cases
belong to the thick Galactic disk (with 
$\rm{[Fe/H]} \approx> -1.0$). The catalog from 
Venn~et\,al.~(2004) lists metallicities,
relative titanium and yttrium abundances for
781 field stars spanning a wide range of metallicities
coincident with the metallicity range for field 
RR~Lyraes. We adopted the relative abundances of the elements
considered in 435 Cepheids from Luck (2018).

\section {ERRORS OF THE DETERMINATION
OF RELATIVE ABUNDANCES OF CHEMICAL ELEMENTS}

In the spectra of RR~Lyraes -- periodic pulsating
variables -- the most symmetric and sharp lines, which
are best suited for elemental abundance determinations,
form near the phase of minimum light and maximum
radius of the star ($\varphi \approx 0.35$), when shocks do not
propagate across the atmosphere (e.\,g., Sneden et\,al.,~2011). 
Most of the spectra used for analyzing the lines
of the elements considered in the original studies were
acquired near the quiescent phase of minimum light.
However, the relative abundances of most of the
chemical elements, including those studied in this
paper, are shown to be practically independent of
phase (unlike the temperature and surface gravity) (see
Pancino et\,al.,~2015, and references therein).

Table 2 gives the statistical information about the
abundances of the elements considered in this paper
and their errors. The second column gives the number
of RR~Lyraes with known relative abundances of the
corresponding element. The next column gives the
mean error computed from the uncertainties quoted in
the original papers. The fourth column gives the standard
deviation of the computed weighted mean relative
abundances for the cases of two or more determinations
of the abundance of each element for a particular
star. This standard deviation provides an estimate
for the external agreement of [el/Fe] determinations
by different authors. As is evident from the Table, the
standard deviation of the computed weighted average
values is smaller than the average errors quoted by the
authors of the original papers. We plot the histograms
of the deviations of published abundances for scandium,
titanium, and yttrium from the corresponding
computed weighted averages in Fig.~1. Unfortunately,
the number of overlapping determinations is small,
especially for yttrium. The resulting distributions are
single-peaked and can be described by the normal law,
and this fact allows us to view errors as random. Hence
our computed weighted average abundances of scandium,
titanium, and yttrium and the single determinations
of relative abundances of these elements compiled
from original papers can be considered reliable
and can be used to analyze the behavior of [el/Fe] in
RR~Lyrae type variables.

\section {COMPARATIVE ANALYSIS
OF THE RELATIVE ABUNDANCES OF SC, TI,
AND Y IN METAL--RICH FIELD STARS
OF VARIOUS TYPES}

To clarify the causes of anomalously low [Sc/Fe],
[Ti/Fe], and [Y/Fe] abundance ratios, we compare
the abundances of the said elements in metal-rich 
RR~Lyraes and other groups of stars, both stationary and
variable. This will also allow us to see how justified are
the explanations of the behavior of [el/Fe] of the
selected elements in the above papers. Fig.~2 shows the
positions of RR~Lyrae type variables (the large red circles)
on the [Fe/H] -– [el/Fe] diagrams for scandium
(panels a,d, and g), titanium (panels b,e, and h), and
yttrium (panels c,f, and i). The vertical dashed line
drawn through $\rm{[Fe/H]} = -1.0$, subdivides stars into
two groups by metallicity. As is evident from the figure,
in all panels in Fig.~2 RR~Lyraes are located along
the strip, which in the domain $\rm{[Fe/H]} > -1.0$ goes
below the domains occupied by the comparison stars
(see below). A clear dependence of [Sc/Fe], [Ti/Fe],
and [Y/Fe] on [Fe/H] is immediately apparent so that the 
abundances of all the elements considered decrease with
increasing metallicity.

The positions of comparison stars are shown in all
panels in Fig.~2. In Figs.~2a,~2b,~2c we added
7066 nearby dwarfs, subgiants, and turnoff stars of the
disk subsystems from Buder et\,al.~(2019). All stars of
this large catalog except four stars with known
[Sc/Fe], ten stars with [Ti/Fe], and ten stars with
[Y/Fe], have $\rm{[Fe/H]} > -1.0$. The large number of
metal-rich stars is what makes it convenient to compare
the behavior of metal-rich RR~Lyraes with the
stars of this catalog. Most of the metal-rich RR Lyraes
in the [Fe/H]--[Sc/Fe] (Fig.~2a) and [Fe/H]--[Y/Fe]
(Fig.~2c) diagrams lie below the domain occupied by
stars from Buder et\,al.~(2019). The only exceptions are
two stars -- KP Cyg and TV Lib, which lie inside the
domain of comparison stars. However, we already
wrote about the peculiarities of the abundances of
other elements and about the anomalously high velocity
of the rotation of these stars about the Galactic
center in our earlier paper (Marsakov et\,al.,~2018a).
Metal-rich RR~Lyraes in the ``[Ti/Fe] -- [Fe/H]'' diagram
(Fig.~2b) lie below the comparison stars, and
only four metal-rich RR~Lyraes are located inside the
compact region occupied by stationary field stars.
Note scandium and yttrium abundances are determined
only from lines of ionized Sc II and YII atoms,
whereas the titanium abundances are in most of the
cases determined by averaging the measurements
based on lines of neutral and ionized titanium (Ti I
and Ti II) (see discussion below).

Figures~2d,~2e,~2f shows the refined positions of the
[el/Fe] ratios of stars and field RR~Lyraes in the low
metallicity domain. In Fig.~2d we had to plot, for comparison,
the [Sc/Fe] ratios for 171 nearby F-G-type
dwarfs from catalog by Reddy et\,al.~(2006), which
include only 15 metal-poor objects. On the other
hand, Figs.~2e,~2f shows 781 field stars from Venn et al.
(2004) spanning the same metallicity range as field
RR~Lyraes (we could not find scandium abundance
determinations by these authors). We can see that in
the $\rm{[Fe/H]} < -1.0$ interval the loci of RR Lyraes and
comparison stars coincide in panels (d--f) of Fig.~2.
On the other hand, at higher metallicities, like in the
case of the comparison with the data from Buder et\,al.~(2019), 
the relative scandium, titanium, and yttrium
abundances in most of the RR~Lyraes are lower than in
stationary comparison stars beyond the quoted errors.

It was repeatedly argued that anomalously low
abundances of some elements in metal-rich RR~Lyraes
are found because stationary stars are used for comparison.
Such peculiarities were believed to be due to
pulsations of the atmospheres of variable stars. Let us
now see how the [Sc/Fe], [Ti/Fe], and [Y/Fe] ratios
depend on metallicity for RR Lyraes compared to similar
dependencies for Cepheids, which represent a different
type of radially pulsating variables. Panels (g--i)
in Fig.~2 show 435 Cepheids from Luck~(2018). All
these Cepheids (except one) are metal rich and populate
rather compact regions on the diagrams. As is evident
from the figures, Cepheids are located significantly
above the strip of metal-rich RR~Lyraes (this is
especially apparent for scandium and yttrium). The
average relative abundances for Cepheids are
$\langle\rm{[Sc/Fe]_{ceph}}\rangle = 0.33 \pm 0.01$, 
$\langle\rm{[Ti/Fe]_{ceph}}\rangle = 0.13 \pm 0.01$, 
$\langle\rm{[Y/Fe]_{ceph}}\rangle = 0.20 \pm 0.01$, whereas the average
relative abundances of the same elements in stationary
comparison stars are either equal to the corresponding
solar values or higher by 0.1 dex, whereas in the metallicity
interval occupied by Cepheids ($\rm{[Fe/H]} > -1.0$)
the relative abundances in field RR Lyraes are equal to
$\langle\rm{[Sc/Fe]_{RRL}}\rangle = -0.34 \pm 0.04$, 
$\langle\rm{[Ti/Fe]_{RRL}}\rangle =-0.08 \pm 0.04$, 
$\langle\rm{[Y/Fe]_{RRL}}\rangle = -0.44 \pm 0.06$. It is clear
that the differences between the abundances in RR
Lyrae type variables with $\rm{[Fe/H]} > -1.0$ and the corresponding
abundances in Cepheids far exceed observation
errors. We do not believe that such low relative
abundances of the elements considered in metal-rich
RR~Lyraes could be due to the fact that these stars are
pulsating variables.

\section {INFLUENCE OF NLTE EFFECTS
ON THE DETERMINATION
OF ELEMENTAL ABUNDANCES} 

It is conceivable that the differences between abundances
of metal-rich RR Lyraes and comparison stars
could be due to nLTE effects. Thus Asplund (2005)
argue that giants are more subject to deviations from
LTE than, e.\,g., dwarfs. However, our comparison stars
also include giants, and Cepheid surface gravities are
often weaker than those of RR Lyraes. Note that the
[Sc/Fe], [Ti/Fe], and [Y/Fe] rations in Cepheids are
much higher than in RR Lyraes of similar metallicities.

Deviations from LTE are also believed to have
stronger effect on lines of neutral atoms than on ion
lines (e.\,g., Bergemann,~2011; Hansen et\,al.,~2011). The
authors of the papers that served as data sources for
relative abundances [el/Fe] used only lines of ionized
elements Sc II and YII to determine the scandium and
yttrium abundances, and computed the [Ti/Fe] ratios
using both neutral and ionized titanium lines (Ti I and
Ti II). The authors of most of the original papers
determined the [Ti/Fe] ratios by as averages of
[Ti I/Fe] and [Ti II/Fe]. We used a similar procedure
in the cases where the paper provided separate [Ti
I/Fe] and [Ti II/Fe] measurements and then computed
the weighted average of the data provided by different
sources. An analysis of the data from the papers
used in this study shows constant offset between
[Ti I/Fe] and [Ti II/Fe], so that the abundances
inferred from Ti I are usually higher than those
inferred from Ti II. For and Sneden~(2010); For et al.
(2011); Pancino et\,al.,~(2015) obtained a similar result.
For and Sneden~(2010); For et\,al.,~(2011) suggested that
such an offset is likely to be due to the uncertainty of
log\,gf values for Ti I lines. Bergemann~(2011) strongly
suggest that [Ti/Fe] ratios for giants should be computed
using solely the data for Ti II lines, however,
they also point out that nLTE effects on the Ti I abundance
shows up mostly in stars with low metal abundance.
Our sample of metal-rich RR~Lyraes include
cases of slight differences between the relative abundances
of Ti I and Ti II, which are mostly within the
errors. Note that the above pattern persists even if we
use only [Ti II/Fe] abundance ratios.

Apparently, the uniqueness of the chemical composition
of metal-rich RR~Lyraes cannot be explained
by nLTE effects.

\section {EUROPIUM, ZIRCONIUM,
AND LANTHANUM ABUNDANCES}

As is well known, SNe II eject $r$-process elements
and, in particular, europium, along with $\alpha$-elements.
That is why the relative abundances of this easily measured
element behave like $\alpha$-elements as a function of
metallicity, allowing the behavior of the [el/Fe]--
[Fe/H] dependence to be more reliably determined
from only one type of chemical elements. Unfortunately,
no europium or other $r$-element abundances
could be found for any field RR~Lyrae with $\rm{[Fe/H]} > -1.0$. 
At the same time, europium abundances are
reliably determined for other metal-rich variablesЎЄ
Cepheids. The [Eu/Fe] ratios in Cepheids are the
same as in field dwarfs and giants (Marsakov et\,al.,~2013). 
What is the case of the lack of data about
europium abundances in metal-rich RR~Lyraes? Perhaps
the authors of original papers did not attempt to
determine these abundances. However, it might be
possible that the europium abundance in metal-rich
RR~Lyraes is too small for its abundances to be determined.
Note that $r$-process elements form in type II
supernovas with masses $8-10M_{\odot}$, whereas more massive
supernovas mostly eject $\alpha$-elements (see, e.\,g.,
Woosley et\,al.,~1994). We can therefore assume that the
matter from which metal-rich RR~Lyraes formed was
enriched mostly by massive SNe II.

We also pointed out the lack of the abundances of
such $s$-elements as lanthanum and zirconium in metal-rich
RR~Lyraes in our catalog (Marsakov et\,al.,~2018a).
On the other hand, La and Zr abundances have been
determined in equally metal-rich Cepheid variables
and proved to be high. Note that europium, lanthanum,
and zirconium abundances are known for some
metal-poor RR~Lyraes.

Further studies are needed to clarify the cause of
the lack of the determinations of the abundances of
these elements in metal-rich RR~Lyraes.

\section {CONCLUSIONS} 

Thus the anomalously low relative abundances of
scandium, titanium, and yttrium in metal-rich Galactic
RR Lyraes cannot be explained by the errors of
their determination. These low abundances can neither
be explained by effects arising in non-stationary
atmospheres of the stars considered. Such low abundances
are unlikely to be due to effects associated with
deviations from local thermodynamical equilibrium.
We also do not understand the lack of data about europium,
lanthanum, and zirconium abundances in
metal-rich RR~Lyraes.

It appears that RR~Lyraes inherited the aforementioned
abundances of the elements considered from
the parent interstellar matter. Earlier we showed that
low residual velocities and close-to-solar chemical
composition of metal-rich RR~Lyraes indicate that
they belong to the subsystem of the thin Galactic disk.
This implies such a small age of these stars that they
could have reached the horizontal branch only in the
case of high initial helium abundances (Marsakov
et\,al.,~2018a). Based on the above, it is conceivable
that metal-rich field RR~Lyraes must have formed
from interstellar matter with the chemical evolution
history different from that of the parent interstellar
matter of most of the field stars in the solar vicinity. We
suggested in our previous papers that such helium-rich
stars could have come to the solar neighborhood as a
result of radial migration from central regions of our
Galaxy, where such stars have already been found
(Marsakov et\,al.,~2019a), or as a result of the capture of
a massive companion dwarf galaxy during early stages
of the evolution of our Galaxy (Marsakov et\,al.,~2020).

\section*{ACKNOWLEDGMENTS}

We are grateful to T.V. Mishenina for reading the manuscript
and valuable advice.

\section*{FUNDING}

The research was financially supported by the Southern
Federal University, 2020 (Ministry of Science and Higher
Education of the Russian Federation).

\section*{CONFLICT OF INTEREST}

The authors declare no conflict of interest.

\renewcommand{\refname}{REFERENCES}

\newpage

\begin{figure*}
\centering
\includegraphics[angle=0,width=0.99\textwidth,clip]{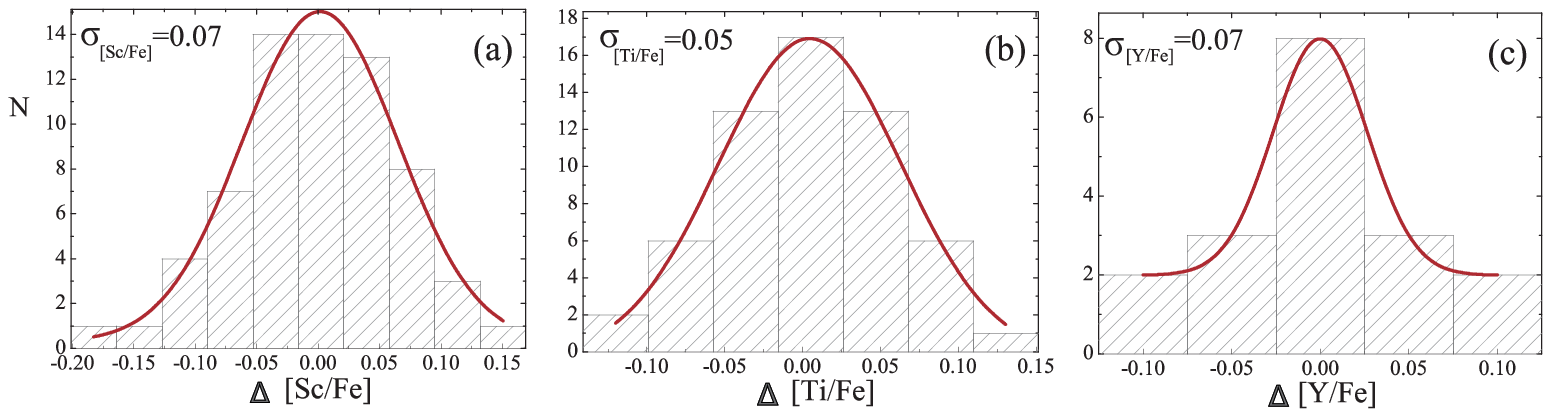}
\caption{Distribution of the deviations $\Delta\rm{[el/Fe]}$ of published
                 values from the corresponding weighted averages for 
                 relative abundances of scandium (a), titanium (b), 
                 and yttrium (c). The histograms are described by 
                 Gaussians. The standard deviations are given on all panels.}
\label{fig1}
\end{figure*}

\newpage

\begin{figure*}
\centering
\includegraphics[angle=0,width=0.99\textwidth,clip]{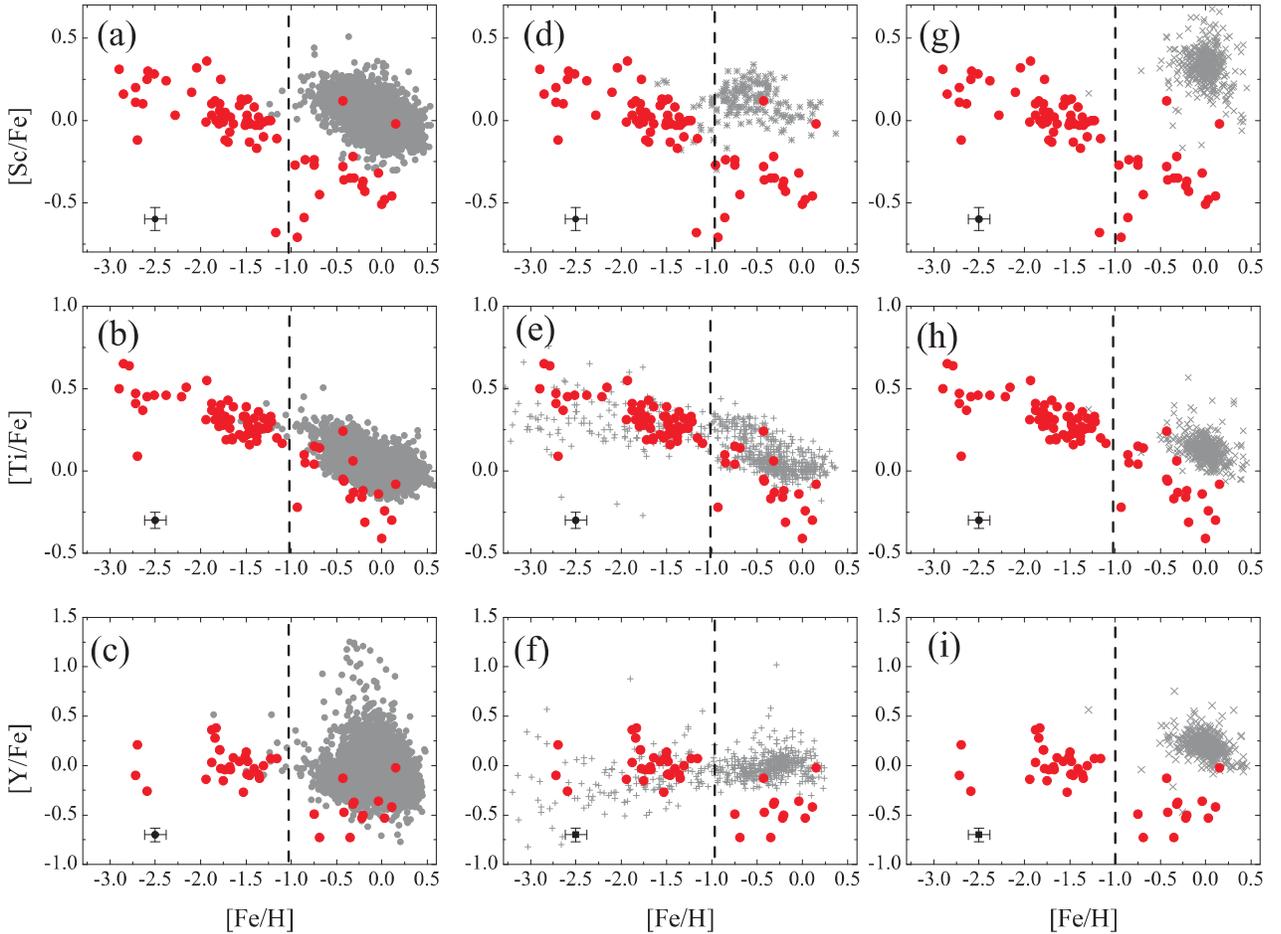}
\caption{Relative abundances of scandium, titanium, and yttrium as
                 a function of metallicity for Galactic-field RR~Lyrae type
                 variables (the large red circles). Comparison stars: 
                 small gray dots-nearby dwarfs, subgiants, and turn-off 
                 stars from Buder et\,al.~(2019) (a--c); gray snowflake
                 symbols-nearby F--G-type dwarfs of the disk subsystems 
                 from Reddy et\,al.~(2006) (d); gray straight crosses-stars
                 of various subsystems from Venn et\,al.~(2004) (e, f);
                 gray slanted crosses--cepheids from Luck (2018) (g--i).
                 Vertical dashed lines separate metal-rich and 
                 metal-poor stars. Also shown are error bars of 
                 weighted average [el/Fe] estimates for RR~Lyraes.}
\label{fig2}
\end{figure*}

\newpage

\begin{table}[t!]
\begin{center}

\caption{%
Metallicities and relative scandium abundances in Galactic-field 
RR~Lyrae type variables (References: 1 -- Andrievsky~et\,al.~(2010),
2 -- Chadid~et\,al.~(2017), 3 -- Clementini~et\,al.~(1995), 
4 -- Clementini~et\,al.~(2000), 5 -- Di Fabrizio~et\,al.~(2002), 
6 -- For and Sneden~(2010), 7 -- For~et\,al.~(2011), 
8 -- Govea~et\,al.~(2014), 9 -- Hansen~et\,al.~(2011), 
10 -- Kolenberg~et\,al.~(2010), 11 -- Liu~et\,al.~(2013), 
12 -- Preston~et\,al.~(2006a), 13 -- Preston~et\,al.~(2006b), 
14 -- Roederer~et\,al.~(2014), 15 -- Sneden~et\,al.~(2017).)}

\bigskip
\label{tproto}

\begin{tabular}{l|c|c|c|c|c|c|c}
\hline \hline
\hspace{-6mm}
& & & & & && \\
\vspace{-6mm}& & & & & & &\\

\multicolumn{1}{c|}{\parbox{1.0cm}{Star}}&
\multicolumn{1}{c|}{\parbox{1.0cm}{[Fe/H], dex}}&
\multicolumn{1}{c|}{\parbox{1.0cm}{[Sc/Fe], dex}}&
\multicolumn{1}{c|}{\parbox{1.5cm}{References [Sc/Fe]}}&
\multicolumn{1}{c|}{\parbox{1.0cm}{Star}}&
\multicolumn{1}{c|}{\parbox{1.0cm}{[Fe/H], dex}}&
\multicolumn{1}{c|}{\parbox{1.0cm}{[Sc/Fe], dex}}&
\multicolumn{1}{c}{\parbox{1.5cm}{References [Sc/Fe]}}\\
\vspace{-3mm} & & & & & & &\\
\hline


\multicolumn{1}{c|}{\parbox{1.0cm}{(1)}}&
\multicolumn{1}{c|}{\parbox{1.0cm}{(2)}}&
\multicolumn{1}{c|}{\parbox{1.0cm}{(3)}}&
\multicolumn{1}{c|}{\parbox{1.0cm}{(4)}}&
\multicolumn{1}{c|}{\parbox{1.0cm}{(1)}}&
\multicolumn{1}{c|}{\parbox{1.0cm}{(2)}}&
\multicolumn{1}{c|}{\parbox{1.0cm}{(3)}}&
\multicolumn{1}{c}{\parbox{1.0cm}{(4)}}\\
\vspace{-3mm} & & & & & & &\\

\hline

              SW And& -0.22& -0.40& 3, 11&     V Ind&-1.45&-0.02& 2\\
              CI And& -0.43& -0.28&    11&    SS Leo&-1.75& 0.03& 2\\
              WY Ant& -1.88&  0.10&  2, 7&    ST Leo&-1.31&-0.10& 2\\
              XZ Aps& -1.79&  0.06&  2, 7&    CM Leo&-1.93& 0.36& 5\\
              BS Aps& -1.48&  0.00&  2, 7&    TV Lib&-0.43& 0.12& 11\\
              AA Aql& -0.32& -0.22&    11&    RR Lyr&-1.49& 0.03& 3,10,11\\
              SW Aqr& -1.38& -0.17&     2&    CN Lyr&-0.04&-0.32& 11\\
              BR Aqr& -0.69& -0.45&    11&    IO Lyr&-1.35&-0.03& 11\\
              DN Aqr& -1.76&  0.01&     2&    KX Lyr&-0.42&-0.36& 11\\
              FV Aqr& -2.59&  0.25&    14&     Z Mic&-1.51&-0.03& 2, 7\\
               X Ari& -2.51&  0.28&  2, 3&    RV Oct&-1.64&-0.02& 2, 7\\
 ASAS J085254-0300.3& -1.53&  0.12& 8, 15&    UV Oct&-1.75&-0.01& 2, 7\\
ASAS J101332-0702.3*& -1.73& -0.12&    15& V 413 Oph&-0.75&-0.27& 11\\
ASAS J132225-2042.3*& -0.96& -0.27&    15& V 445 Oph& 0.11&-0.46& 2, 3, 11\\
ASAS J143322-0418.2*& -1.48& -0.02&    15&    AO Peg&-1.26& 0.00& 11\\
ASAS J203145-2158.7*& -1.17& -0.68&    15&    AV Peg&-0.19&-0.43& 2\\
 ASAS J162158+0244.5& -1.84&  0.12& 8, 15&    DH Peg&-1.31&-0.01& 11\\
              RS Boo& -0.21& -0.37&    11&    HH Pup&-0.93&-0.71& 2\\
              ST Boo& -1.73&  0.05&     3& V 701 Pup&-2.90& 0.31& 8, 15\\
    BPS CS 22881-039& -2.72&  0.20&6,9,13,14&SV Scl*&-2.28& 0.03& 15\\
    BPS CS 22940-070& -1.41&  0.08&  6,14&    VY Ser&-1.78& 0.05& 2,3,11\\
    BPS CS 30317-056& -2.85&  0.16&     9&    AN Ser& 0.00&-0.51& 2\\
    BPS CS 30339-046& -2.70& -0.12&    14& V 456 Ser&-2.64& 0.10& 14\\
              YZ Cap& -1.50& -0.01& 8, 15&     T Sex&-1.55& 0.13& 15\\
              RR Cet& -1.48&  0.13&2,3,11&   RU Sex*& -2.1& 0.17& 15\\
              UU Cet& -1.36&  0.02& 3, 11& V 440 Sgr&-1.16&-0.11& 3, 11\\
              CU Com& -2.38&  0.24&     4&V 1645 Sgr&-1.94&-0.01& 2, 7\\
               W Crt& -0.75& -0.24&     2&   MT Tel*& -2.58& 0.3& 15\\
              Y Crv*& -1.39& -0.03&    15&     W Tuc& -1.76&0.02& 2\\
              DM Cyg&  0.03& -0.48&11&TYC 4887-622-1& -1.79&0.10& 8, 15\\
              KP Cyg&  0.15& -0.02&1&TYC 6644-1306-1& -1.78&0.25& 8, 15\\
              DX Del& -0.31& -0.35&2,11&TYC 8776-1214-1&-2.72&0.11& 8, 15\\
              CS Eri& -1.70& -0.13&    15&    CD Vel& -1.67&0.02& 2, 7\\
              SX For& -1.80& -0.02&     2&    ST Vir& -0.85&-0.24& 2\\
              TY Gru& -1.88&  0.03&  7,12&    UU Vir& -0.86&-0.59& 2\\
              BO Gru& -1.83&  0.00& 8, 15&    AS Vir& -1.68&-0.07& 2, 7\\
              TW Her& -0.35& -0.35&    11&   AU Vir*& -2.04& 0.32& 15\\
              VX Her& -1.46& -0.13&  3,11&    DO Vir& -1.57& 0.09& 11\\
              DT Hya& -1.23&  0.00&  2, 7&       -  &   -  &  -  & - \\

\hline
\end{tabular}
\end{center}

\end{table}

\newpage

\begin{table}[t!]
\begin{center}

\caption{%
Statistics of the relative abundance determinations
and their errors}
\bigskip
\label{tproto}

\begin{tabular}{l|c|c|c}
\hline \hline
\hspace{-6mm}
& & &  \\
\vspace{-6mm}& & & \\

\multicolumn{1}{c|}{\parbox{2.0cm}{Element}}&
\multicolumn{1}{c|}{\parbox{2.0cm}{Number of stars}}&
\multicolumn{1}{c|}{\parbox{2.0cm}{Average quoted error}}&
\multicolumn{1}{c}{\parbox{2.5cm}{Standard deviation of the quoted mean}}\\

\vspace{-3mm} & & & \\

\hline

Fe& 100& 0.14& 0.12 \\
Sc&  77& 0.14& 0.07 \\
Ti&  83& 0.13& 0.05\\
 Y&  43& 0.11& 0.07\\

\hline
\end{tabular}
\end{center}

\end{table}

\newpage

{}
\end{document}